\documentclass[aps,prl,twocolumn,floatfix,superscriptaddress]{revtex4-2}
\usepackage[utf8]{inputenc}
\usepackage[T1]{fontenc}
\usepackage[english]{babel}
\usepackage{mathtools}
\usepackage{newtxtext}  
\usepackage{amsmath}
\usepackage{amsfonts}

\usepackage{booktabs}

\renewcommand\vec{\mathbf}

\usepackage[dvipsnames,svgnames,x11names,hyperref]{xcolor}
\usepackage{hyperref}
\hypersetup{ 
    colorlinks=true, linkcolor=NavyBlue, urlcolor=NavyBlue, citecolor=NavyBlue
}

\usepackage{physics,braket,siunitx,slashed}
\usepackage[capitalise]{cleveref}

\usepackage{tabularx}

\newcommand{\bk}{\mathbf{k}}
\newcommand{\br}{\mathbf{r}}

\newcommand{\bq}{\mathbf{q}}
\newcommand{\bP}{\mathbf{P}}
\usepackage[normalem]{ulem}

\begin{document}

\title{Cooper Instability of a Magnetic Wigner Crystal}

\author{Wangqian Miao}
\affiliation{Department of Physics, The Pennsylvania State University, University Park, Pennsylvania 16802, USA}
\affiliation{Center for Theory of Emergent Quantum Matter,
The Pennsylvania State University, University Park, Pennsylvania 16802, USA}

\author{Chunli Huang}
\affiliation{Department of Physics and Astronomy, University of Kentucky, Lexington, Kentucky 40506-0055, USA}

\date{\today}

\begin{abstract}
Recent scanning tunneling microscopy has directly imaged electronic crystals in rhombohedral hexalayer graphene, and magnetotransport measurements suggest that superconductivity and charge-density-wave order emerge from a common spin-valley-polarized ferromagnetic metal at nearby carrier densities. Motivated by these observations, we ask whether collective modes in a magnetic Wigner crystal can mediate pairing in this setting. We address this question by studying the Cooper problem for two electrons in the first conduction band of the Wigner crystal. Unlike conventional phonon-mediated pairing, the collective mode (boson) energies and single-particle (fermion) dispersion are comparable scales controlled by the electron density, leaving no small parameter for an adiabatic expansion. Moreover, because the magnetic Wigner crystal is not time-reversal invariant, the usual Cooper logarithmic divergence in the pairing susceptibility is absent, so the attraction must exceed a finite threshold for pairing to occur. These unusual features motivate a numerical experiment to see whether the crystal's own vibrations can bind electrons. We obtain the crystalline ground state and its collective mode spectrum within Hartree--Fock and time-dependent Hartree--Fock approximation, and derive the electron--phonon Hamiltonian using the quasiboson approximation. Solving the resulting two-electron problem, we find that phonon-mediated attraction exceeds the pairing threshold, with chiral $p$-wave pairing as the leading instability. Within these approximations, we show that collective modes of a Wigner crystal can indeed favor chiral pairing despite broken time-reversal symmetry. Our results provide an encouraging starting point for developing a full many-body theory to study electronic crystallization and superconductivity.
\end{abstract}

\maketitle

\textcolor{NavyBlue}{\textit{Introduction.}} Interaction-driven states that break translational symmetry have recently joined the rich phase diagram of dilute electrons in rhombohedral graphene~\cite{zhou_half-_2021,zhou_superconductivity_2021,han_orbital_2023,patterson_superconductivity_2025,yang_impact_2025,seo_family_2026,cea_superconductivity_2022,szabo_metals_2022,qin_functional_2023}. These include charge-density waves~\cite{ron_hierarchy_2025,nguyen_coexisting_2026,aronson_displacement_2025}, stripe states~\cite{qin25stripe}, and Wigner crystals~\cite{Wigner_original,han2026WC,zhou_competing_2026,zheng_chiral_2026,sun_real-space_2026,lu_extended_2025,waters_chern_2025,xiang_continuously_2025}, defined here as electronic crystals with one electron per unit cell~\cite{miao2026ahc,ge_nematic_2026,abouelkomsan_rhombohedral_2026,guo2025ahcblg}. The most direct evidence comes from scanning tunneling microscopy of hole-doped rhombohedral hexalayer graphene, where periodic electronic crystals have been directly imaged in real space~\cite{sun_real-space_2026}. Transport measurements provide complementary indications of charge ordering. In electron-doped hexalayer graphene, the carrier density inferred from the Hall response is inconsistent with that expected from the observed quantum Hall states \cite{ron_hierarchy_2025,nguyen_coexisting_2026}. In tetralayer graphene, normal-state quantum oscillation frequencies are difficult to reconcile with those expected for fully polarized annular, nematic, or three-pocket states~\cite{kalantre_fermiology_2026}, motivating scenarios in which a fraction of the carriers is localized~\cite{dong26metalicWC,feng26metalicWC,kim_shape_2026}.

These electronic crystals (or charge density waves) have two unique properties. First, they occur in a low-density regime where the parent metal is already spin-valley-polarized, allowing crystalline order to coexist with broken time-reversal symmetry~\cite{zhou_half-_2021,han_signatures_2025,choi_superconductivity_2025,dutta_reconfigurable_2026,dong_anomalous_2024,soejima_anomalous_2024,dong_theory_2024,zhou24moireless,tan_parent_2024,zeng_sublattice_2024,joy_chiral_2025,zverevich_spin-triplet_2026}. Second, at a dilute filling of order $10^{-4}$ carriers per carbon atom, the spacing between electrons is roughly two orders of magnitude larger than the atomic spacing. These electronic crystals are therefore not pinned by lattice commensuration but by smooth disorder and/or sample boundaries.
Most importantly, transport measurements place chiral superconductivity~\cite{han_signatures_2025,choi_superconductivity_2025,dutta_reconfigurable_2026,zheng_chiral_2026} adjacent to a charge-ordered phase on its higher-density side~\cite{han2026WC,zhou_competing_2026}, while hexalayer measurements reveal intertwined superconductivity and charge order~\cite{ron_hierarchy_2025,nguyen_coexisting_2026}. Theoretical routes to unconventional and topological superconductivity in rhombohedral graphene and doped Chern insulators have also been proposed~\cite{ghazaryan_unconventional_2021,you_kohn-luttinger_2022,geier_chiral_2026,chou_intravalley_2025,parra-martinez_band_2025,yang_topological_2025,christos_finite-momentum_2026,yoon_quarter_2026,kim_topological_2025,jahin_enhanced_2026,shi_doping_2025}. This proximity motivates us to ask whether phonons and other intraflavor collective modes of the electronic crystal can mediate pairing between spin-valley-polarized electrons. Pairing mediated by isospin fluctuations has been studied in other polarization regimes~\cite{chatterjee_inter-valley_2022,dong_superconductivity_2021,dong_spin-canting_2026}. 
The SVP parent metal considered here has only one active flavor at the Fermi surface. A single spin- or valley-flip excitation takes an electron outside this low-energy sector, so pairing mediated by these fluctuations should rely on higher-order processes involving multiple magnons. By contrast, collective modes of nearby Wigner crystal couple directly to charge without changing the spin or valley flavor.

In this work, we test this possibility by solving the Cooper problem~\cite{cooper_bound_1956,karuzin_berry-flux-controlled_2026} for two electrons in the first conduction band of a Wigner crystal. The problem differs from conventional phonon-mediated pairing~\cite{chou_acoustic-phonon-mediated_2021,vinas_bostrom_phonon_2024} in two important respects: the phonon energies are comparable to the electronic bandwidth, and broken time-reversal symmetry removes the usual guarantee of a Cooper instability at arbitrarily weak attraction. Nevertheless, we find that the attraction mediated by Wigner-crystal phonons is strong enough to bind the two electrons, with the leading instability occurring in a chiral $p$-wave channel.

\begin{figure}[t]
\centering
\includegraphics[width=\columnwidth]{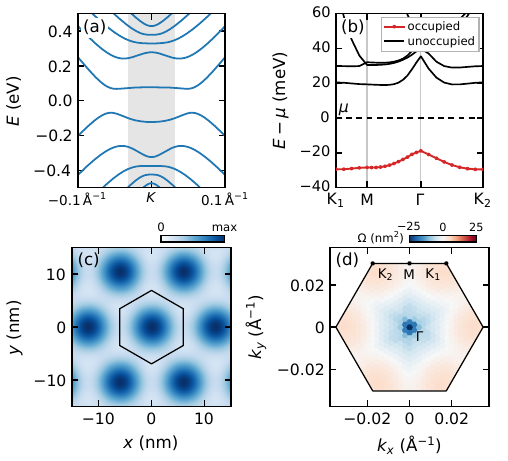}
\caption{(a) Single-particle bands of R5G near the $K$ valley at displacement-field potential $U=$\,$50$~meV. The shaded window marks the momenta $|\delta\bk|\le k_\mathrm{F}$ occupied at the density $n$ in a single flavor.
(b) HF bands of the polarized flavor with the occupied band in red and the chemical potential $\mu$ set at midgap. (c) Charge density of the Wigner crystal; the hexagon marks the Wigner--Seitz cell. (d) Berry curvature $\Omega(\bk)$ of the occupied band; its integral vanishes, $C=0$. 
}
\label{fig1}
\end{figure}

\begin{figure*}[t]
\centering
\includegraphics[width=2.1\columnwidth]{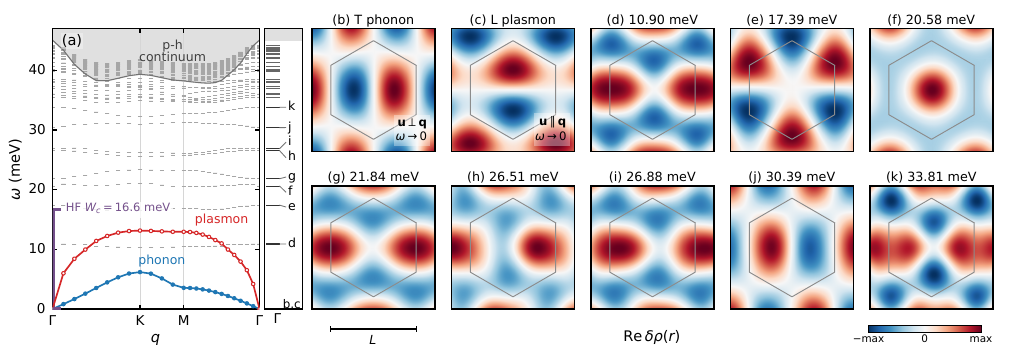}
\caption{Particle-hole excitation spectrum of the magnetic Wigner crystal. Transverse phonon (blue), longitudinal plasmon (red), and uncorrelated HF particle–hole continuum (shaded). The adjacent strip resolves collective modes at $\Gamma$; $W_c=16.6$ meV is the first conduction-band width. (b,c) Transverse and longitudinal modes $\delta\rho=\vec{u}\cdot\nabla\rho$ with $\rho$ shown in Fig.~1c. (d–k) Other collective modes at $\Gamma$. Colors show $\operatorname{Re}\delta\rho(\mathbf r)$, normalized separately for each mode. Hexagons outline the Wigner–Seitz cell. $L$ is the lattice constant.}
\label{fig2}
\end{figure*}

\textcolor{NavyBlue}{\textit{Model.}}
We consider the standard continuum description of the tight-binding Hamiltonian of rhombohedral pentalayer graphene~\cite{herzog-arbeitman_moire_2024},
\begin{align}
H=&\sum_{\bk s\tau}\sum_{\alpha\beta}c^\dagger_{\bk s\tau\alpha}\,[h_\tau(\bk)]_{\alpha\beta}\,c_{\bk s\tau\beta}\nonumber\\
&+\frac{1}{2}\sum_{\bq}\sum_{\ell\ell'}v_{\ell\ell'}(\bq)\,{:}\rho_\ell(\bq)\rho_{\ell'}(-\bq){:},
\label{eq:H}
\end{align}
 Here $c^\dagger_{\bk s\tau\alpha}$ creates an electron with spin $s$ and valley $\tau$ in the orbital $\alpha=(\ell,\sigma)$ on layer $\ell$ and sublattice $\sigma$, with momentum $\bk$ measured from the valley momentum, and $h_\tau(\bk)$ is the Slater--Koster Hamiltonian of valley $\tau$ including the displacement field [Fig.~\ref{fig1}(a)].
 The layer-projected charge density is $\rho_\ell(\bq)=\sum_{\bk s\tau\sigma}c^\dagger_{\bk+\bq,s\tau\ell\sigma}c_{\bk,s\tau\ell\sigma}$.
 The electrons interact through the dual-gate-screened, layer-resolved Coulomb potential $v_{\ell\ell'}(\bq)=e^2\tanh(qd_{\rm gate})\,e^{-qd_0|\ell-\ell'|}/(2S\varepsilon\varepsilon_0q)$, where $S$ is the sample area, $d_0=0.335$nm, $\varepsilon=5$, $d_{\rm gate}=40$~nm. We work with an electron density $n=0.8\times10^{12}$~cm$^{-2}$ close to the metallic Wigner crystal regime in our previous calculations \cite{miao2026ahc,miao_retarded_2026}, and a displacement-field potential $U=$\,$50$~meV between adjacent layers; the remaining single-particle terms are listed in the SM~\cite{Supp}.
 
 Starting from this purely repulsive model, we construct an effective fermion--quasiboson Hamiltonian through a sequence of approximations and solve the resulting two-electron Cooper problem. First, we use the Hartree--Fock approximation (HFA)~\cite{Needs_HFWC} to obtain a fully spin-valley-polarized Wigner crystal. We refer to this state as a magnetic Wigner crystal because it breaks both translational and time-reversal symmetries. Only one spin-valley flavor is occupied, chosen spontaneously in our model, and the other three flavors are unoccupied. Second, we use time-dependent Hartree--Fock approximation (TDHFA) to calculate the crystal's collective modes and found many collective modes in addition to the transverse phonon and plasmon. Within the quasiboson approximation, we treat these collective excitations as bosons and obtain their coupling to the electrons. Finally, we solve the Cooper problem for two electrons in the first conduction band of the magnetic Wigner crystal to determine whether phonon exchange produces a bound pair.


\textcolor{NavyBlue}{\textit{HFA and TDHFA.}}
The lattice constant of a Wigner crystal with one electron per unit cell, $L=[2/(\sqrt{3}n)]^{1/2}$,
is fixed by the density $n$. 
In HFA, the Bloch wavefunction $\psi_{n \bk}$ satisfies the eigenvalue equation 
\begin{equation}
    [h_0(\bk)+\Sigma^{\rm HF}(\bk)]\psi_{n \bk}=\epsilon_{n\bk}\psi_{n\bk}
\end{equation}
where $h_0(\bk)$ is the first conduction band (Fig.~1a) and $\Sigma^{\rm HF}(\bk)$ is the self-energy to be determined self-consistently. The momentum $\bk$ sits in the Wigner-crystal Brillouin zone, which we discretize into a $24\times24$ momentum grid for most calculations. The lowest-energy solution we find is fully spin and valley polarized, with one occupied band separated from the empty bands by an energy gap of roughly $38$~meV [Fig.~\ref{fig1}(b)]. 
This gap is of the same order as the characteristic Coulomb scale $E_C=e^2/(4\pi\varepsilon_0\varepsilon L)\simeq24$~meV.
Its charge density forms a triangular Wigner crystal [Fig.~\ref{fig1}(c)], and the occupied band carries a Berry curvature that integrates to zero in the mBZ [Fig.~\ref{fig1}(d)].


Next, we diagonalize the generalized-RPA eigenvalue equation  to compute the intra-flavor particle-hole excitation spectrum~\cite{cote1991collective,miao_retarded_2026, kwan_moire_2025,ambuj2025phonon},
\begin{equation}
\begin{pmatrix}A_{\bq}&B_{\bq}\\-B^*_{-\bq}&-A^*_{-\bq}\end{pmatrix}
\begin{pmatrix}X^\nu_{\bq}\\Y^\nu_{-\bq}\end{pmatrix}
=\Omega_\nu(\bq)\begin{pmatrix}X^\nu_{\bq}\\Y^\nu_{-\bq}\end{pmatrix},
\label{eq:grpa}
\end{equation}
where $A$ and $B$ are $N_kN_b\times N_kN_b$ matrices, with $N_k$ the number of $\bk$ points and $N_b$ the number of unoccupied HF bands~\cite{Supp}. 

\begin{figure}[b]
\centering
\includegraphics[width=\columnwidth]{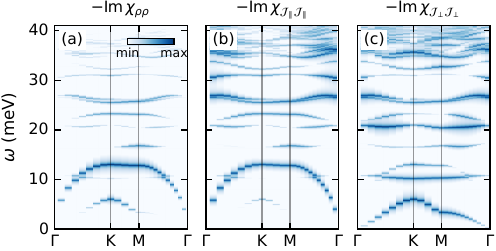}
\caption{Spectral density of the magnetic Wigner crystal  for density response $-{\rm Im}\,\chi_{\rho\rho}$, longitudinal-current response $-{\rm Im}\,\chi_{\mathcal{J}_\parallel\mathcal{J}_\parallel}$, and transverse-current response $-{\rm Im}\,\chi_{\mathcal{J}_\perp\mathcal{J}_\perp}$. The collective modes above plasmon and transverse phonon take up a lot of spectral weight.
Each map uses Lorentzian broadening of $0.065$~meV, is normalized to its own maximum, and is shown on a logarithmic scale over three decades. }
\label{fig3}
\end{figure}

The excitation spectrum is shown in Fig.~2a. Two modes become gapless as $q\rightarrow0$: the longitudinal plasmon and the transverse sound mode. Both are described by elasticity theory in terms of a displacement field $\vec u(\vec r,t)$, which translates the converged HF charge density $\rho\rightarrow \rho'=\rho+  \vec u\cdot \nabla \rho$. In the two-dimensional electron gas, the TDHF dispersions of these two modes agree well with classical calculations based on delta-function point charges already $r_s=8$, even though the equilibrium density remains broadly distributed around each lattice site~\cite{ambuj2025phonon}. 
The gray region marks the uncorrelated particle--hole continuum, whose lower edge is $
E_{\rm ph}^{\rm HF}(\bq)
=\min_{n>0,\bk}\left[\epsilon_{n,\bk+\bq}-\epsilon_{0,\bk}\right]$.
In TDHFA, finite Coulomb matrix interaction between particle--hole pairs shift $E_{\rm ph}^{\rm HF}(\bq)$ downward in energy. Beyond the two translational modes \cite{dong_anomalous_2024}, the TDHF spectrum contains several isolated collective modes below the particle--hole continuum. Fig.~\ref{fig2}(b)--(k) show the snapshot of the density fluctuations associated with these modes, with red and blue indicating electron accumulation and depletion. Such internal modes are not uncommon in quantum solids, e.g.~solid $^4$He~\cite{gazit2016collective}.
A low-energy theory describing them can be constructed by expanding in slowly varying fluctuations of the magnitudes and relative phases of the Wigner crystal's dominant Fourier components.   Fig.~\ref{fig3} shows that they carry a lot of spectral weight in the density--density and current--current response functions and can leave strong fingerprints in optical experiments.



\textcolor{NavyBlue}{\textit{Quasiboson approximation.}}
We now use the quasiboson approximation to approximate the particle-hole excitations (fermion bilinears) as bosons and construct an effective fermion-boson Hamiltonian. This approach is motivated by the particle-vibration coupling in nuclear theory ~\cite{rowe2010nuclear,barranco_surface_1999}, where the exchange of collective vibrations generates an induced interaction between nucleons and contributes to pairing correlations, see also~\cite{vasiliou_electrons_2024}.  Although the formalism is general, our aim here is to study pairing between electrons in the first HF conduction band. The fermion-boson Hamiltonian is
\begin{equation}
H_{\rm QB} =
\sum_{\bk}\xi_{\bk}d^\dagger_{c\bk}d_{c\bk}
+\sum_{\nu\bq}\Omega_\nu(\bq)Q^\dagger_\nu(\bq)Q_\nu(\bq)
+H_{fb}.
\end{equation}
Here $\xi_\bk = \epsilon_{c\bk}-\mu$, $d^\dagger_{c\bk}$ creates a fermion in the first HF conduction band $c$, and $Q^\dagger_\nu(\bq)$ creates a boson in mode $\nu$ with momentum $\bq$ and energy $\Omega_\nu(\bq)$. Their coupling is 
\begin{equation}
H_{fb}=\sum_{\nu\bk\bq}g_\nu(\bk,\bq)\,d^\dagger_{c\bk}d_{c\bk+\bq}Q^\dagger_\nu(\bq)+{\rm h.c.},
\label{eq:Hfb}
\end{equation}
\begin{align}
g_\nu(\bk,\bq)=\sum_{ph\bk'}\Big[&\bar{\mathcal V}_{p\bk'+\bq,c\bk;h\bk',c\bk+\bq}\,X^{\nu*}_{ph\bk'}(\bq)\nonumber\\
&+\bar{\mathcal V}_{h\bk',c\bk;p\bk'-\bq,c\bk+\bq}\,Y^{\nu*}_{ph\bk'}(-\bq)\Big],
\label{eq:g}
\end{align}
where $p$ and $h$ label the empty and occupied HF bands and $\bar{\mathcal V}_{12;34}=\mathcal V_{12;34}-\mathcal V_{12;43}$
 is the antisymmetrized matrix element of the Coulomb interaction between HF states.

We now add two electrons to the first conduction band of the Wigner crystal. A single added electron will be dressed by the crystal's phonons, forming a polaron whose excitation energy is generically lower than the HF value, which we estimate in the SM in the configuration space that contains 1 fermion and 1 boson. If this dressing leads to localization, it can also induce a local distortion of the crystal density. With two added electrons, the same electron--phonon coupling generates an induced interaction through virtual phonon exchange. 

Unlike conventional electron-phonon systems, where the large electron--ion mass difference provides a separation of energy scales, the fermions and collective modes here arise from the same underlying electronic degrees of freedom (Eq.~\eqref{eq:H}). The bosons are really fermion bilinears approximated as quasibosons. Moreover, their energies are comparable to the electronic bandwidth, so there is no clear adiabatic small parameter to guide the calculation. Thus, we follow Occam's razor and use the simplest induced interaction, obtained by eliminating virtual one-phonon states to second order in the electron--phonon coupling. We neglect retardation and solve the resulting static two-electron problem. For two electrons in the first conduction band with total momentum $\bP$, we expand their wavefunction in terms of HF orbitals,
\begin{align}
\Psi_{\bP}(x_1,x_2)
=\frac{1}{\sqrt{2}}\sum_{\bk}\chi_{\bP}(\bk)
\Big[
&\psi_{c\bk}(x_1)\psi_{c,\bP-\bk}(x_2)
\nonumber\\
-&\psi_{c\bk}(x_2)\psi_{c,\bP-\bk}(x_1)
\Big].
\label{eq:Psi}
\end{align}
where $x_i=(\br_i,\alpha_i)$ is the coordinate and sublattice component of electron $i$, and $\chi_{\bP}(\bk)$ is the pair amplitude. 
The antisymmetrized form enforces Fermi statistics for electrons of the same flavor. The pair amplitudes satisfy the Schrödinger equation \cite{baym2018lectures}:
\begin{equation}
(\xi_{\bk}+\xi_{\bP-\bk})\chi_{\bP}(\bk)
+{\sum_{\bk'}} \bar{V}_{\bk,\bP-\bk;\,\bk',\bP-\bk'}
\chi_{\bP}(\bk')
=E_{\bP} \chi_{\bP}(\bk),
\label{eq:cooper}
\end{equation}
where $\bar{V}_{\bk,\bP-\bk;\,\bk',\bP-\bk'}=\left\langle \psi_{c\bk},\psi_{c,-\bk+\bq}\right|
V
\left|\overline{\psi_{c\bk'},\psi_{c,-\bk'+\bq}}\right\rangle$ is the antisymmetrize matrix element, $\left|\overline{ij}\right\rangle=\left|ij\right\rangle-\left|ji\right\rangle$.
The quasi-boson-mediated interaction is
\begin{align}
V^{\rm ph}_{\bk,\bP-\bk;\,\bk',\bP-\bk'}
={}&-\sum_\nu^{\nu_{\rm cut}}
\frac{g_\nu(\bk,\bq)\,g_\nu^*(\bP-\bk',\bq)}
{\Omega_\nu(\bq)}
\nonumber\\
&-\sum_\nu^{\nu_{\rm cut}}
\frac{g_\nu(\bP-\bk,-\bq)\,g_\nu^*(\bk',-\bq)}
{\Omega_\nu(-\bq)}.
\label{eq:VD}
\end{align}
with $\vec q= \vec k -\vec k'$.
This matrix element describes scattering from an incoming pair $(\bk',\bP-\bk')$ to an outgoing pair $(\bk,\bP-\bk)$ mediated by a boson with energy $\Omega_\nu$.  The two terms correspond to the two possible time orderings of virtual boson exchange: either electron can emit the phonon, which is then absorbed by the other. 

\begin{figure}[t]
\centering
\includegraphics[width=\columnwidth]{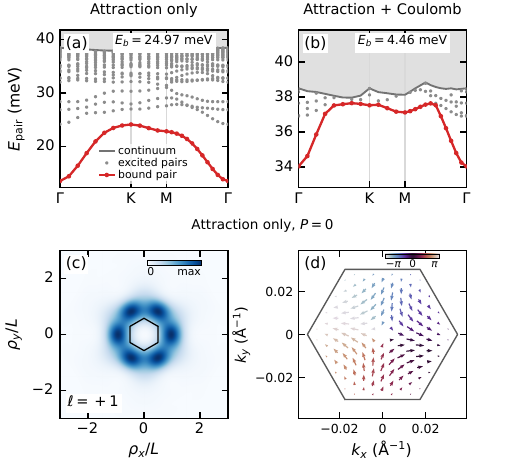}
\caption{Cooper pair in magnetic wigner crystal. (a),(b) Lowest pair energy $E^{\rm pair}_{\bP}$ (red dots), other bound states (gray dots), and the two-particle continuum (shaded), for attraction only and attraction plus residual Coulomb interaction, respectively. The binding energy at $\bP=0$ is indicated. (c),(d) Attraction-only ground state at $\bP=0$. (c) Pair density $\sum_{\alpha\beta}|\Psi_{0}(\vec \rho,\alpha;-\vec \rho,\beta)|^2$ vs relative coordinate $\rho$ with the center of mass fixed at a $C_3$ center of the crystal; the hexagon marks the Wigner--Seitz cell. (d) Winding
$\chi_{\mathbf0}(\bk)$ is 1.}
\label{fig4}
\end{figure}

Fig.~4(a) shows the two-electron spectrum obtained from Eq.\eqref{eq:cooper} with the quasiboson-mediated interaction, retaining modes up to $\nu_{cut}=39$. The lowest pair branch lies below the two-electron continuum throughout the plotted momentum path and has its minimum at $\mathbf P=0$. We define the binding energy as
\begin{equation}
    E_b(\mathbf P)=E_{\mathbf P}^{\mathrm{cont}}-E_{\mathbf P}^{\mathrm{pair}},
\;
E_{\mathbf P}^{\mathrm{cont}}
=\min_{\mathbf k}\left(\xi_{\mathbf k}+\xi_{\mathbf P-\mathbf k}\right)
\end{equation}
where $E_{\mathbf P}^{\mathrm{pair}}$ is the lowest eigenvalue of Eq.~\eqref{eq:cooper}. At $\mathbf P=0$, the induced interaction yields a binding energy of $24.97$ meV.
The corresponding real-space pair density is shown in Fig. 4(c), with the center of mass fixed at a threefold rotation center of the crystal. The density is concentrated at finite relative separation and extends beyond the Wigner–Seitz cell, indicating a pair size comparable to the crystal lattice constant. The momentum-space phase texture of $\chi_0(\mathbf k)$ in Fig.~4(d) exhibits a winding of $\ell=+1$, consistent with chiral $p$-wave pairing.

\begin{table}[b]
\caption{$\Gamma$-point binding energies $E_b(\Gamma)$ (meV) for different sets of retained collective modes. The phonon and plasmon are tracked by mode character across branch crossings.}
\label{tab:mode-binding}
\begin{ruledtabular}
\begin{tabular}{lcc}
Modes retained $\nu_{cut}$ & \shortstack{Attraction\\only} & \shortstack{Attraction $+$\\Coulomb} \\
Phonon $+$ plasmon & 23.72 & 3.57 \\
$\nu_{\rm cut}=10$  & 24.64 & 4.17 \\
$\nu_{\rm cut}=39$ & 24.97 & 4.46 \\
\end{tabular}
\end{ruledtabular}
\end{table}

To test whether the induced attraction overcomes direct Coulomb repulsion, $V^C$, we add 
$\left\langle \psi_{c\bk},\psi_{c,-\bk+\bq}\right|
V^C
\left|\psi_{c\bk'},\psi_{c,-\bk'+\bq}\right\rangle$ to $V^{\rm ph}$ in Eq.~\eqref{eq:VD} and solve the Schrödinger equation again. As shown in Fig.~4(b), Coulomb repulsion substantially weakens the binding, reducing $E_b(0)$ to $4.46$ meV. Nevertheless, the lowest pair branch remains below the continuum along the plotted path, with its minimum at $\mathbf P=0$. Table I compares the contributions of different collective modes to pairing. The two translational modes, the transverse phonon and longitudinal plasmon, provide most of the binding. The internal modes give only a modest enhancement. By $\nu_{\mathrm{cut}}=39$, the retained modes extend into the particle–hole continuum. Despite their large number, these additional modes contribute little because they couple weakly to the conduction-band electrons.

\textcolor{NavyBlue}{\textit{Summary.}}
In this work, we use a series of approximations to show that, starting from a multicomponent electron-gas model appropriate for rhombohedral graphene with purely repulsive Coulomb interactions, electrons can organize into a magnetic Wigner crystal, and two additional electrons in its first conduction band can form a chiral $p$-wave bound state. The relative wavefunction extends over a region slightly larger than the Wigner-crystal unit cell. The picture we have in mind is similar to that of Bardeen--Pines. Here, however, the crystal is a quantum solid with spatially extended density peaks. An electron moving through it can excite not only the two translational modes but also the internal modes shown in Fig.~2. A second electron with opposite momentum can then experience an effective attraction mediated by these internal motions. 

For the Wigner crystal studied here, pairing is dominated by the two translational modes. This is because the conduction-band width is $W_c=16.6$~meV, and only three collective branches lie below this energy scale (Fig.~2). Most internal modes occur at higher energies, reducing their contributions through the energy denominators. A larger conduction-band width could make more internal modes relevant to pairing.

A quantitative treatment of the dynamical picture requires accounting for the relative time scales of electronic motion and collective-mode dynamics, which is quite challenging because more electrons can also modify the crystal itself and its excitation spectrum. We leave this for future work.

\textcolor{NavyBlue}{\textit{Acknowledgement.}} We acknowledge useful discussions with Cyprian Lewandowski and Sandeep Joy. W.M. acknowledges related collaborations with Mu-Yang Chen, Binghai Yan and Chao-Xing Liu together with a fellowship from the Center for Theory of Emergent Quantum Matter (C-TEQ) at Penn State.  C.H. was supported in part by the U.S. Department of Energy, Office of Science, Office of Basic Energy Sciences, under Award No.~DE-SC0024346 and in part by the National Science Foundation CAREER Award No.~DMR-2541471. This work was performed in part at the Aspen Center for Physics, which is supported by a grant from the Simons Foundation (1161654, Troyer) and was supported in part by grant NSF PHY-2309135 to the Kavli Institute for Theoretical Physics (KITP).

\bibliographystyle{apsrev4-2}
\bibliography{reference}

\end{document}